# False Analog Data Injection Attack Towards Topology Errors: Formulation and Feasibility Analysis

Yuqi Zhou, Jorge Cisneros-Saldana, Le Xie
Department of Electrical and Computer Engineering
Texas A&M University
College Station, USA

*Abstract*—In this paper, we propose a class of false analog data injection attack that can misguide the system as if topology errors had occurred. By utilizing the measurement redundancy with respect to the state variables, the adversary who knows the system configuration is shown to be capable of computing the corresponding measurement value with the intentionally misguided topology. The attack is designed such that the state as well as residue distribution after state estimation will converge to those in the system with a topology error. It is shown that the attack can be launched even if the attacker is constrained to some specific meters. The attack is detrimental to the system since manipulation of analog data will lead to a forged digital topology status, and the state after the error is identified and modified will be significantly biased with the intended wrong topology. The feasibility of the proposed attack is demonstrated with an IEEE 14-bus system.

## I. Introduction

In modern power systems, topology processor (TP) performs analysis on the status of circuit breakers (CBs) to determine the system model, which can be further used in state estimation (SE). The status of CBs is correct and known most of the time, but in some cases the assumed status may turn out to be erroneous [1]. This happens when isolation switches are not telemetered or operated. Other possible causes may include unreported breaker manipulation, communication failure, cyber attacks, etc. In these cases, the topology model given by TP will be an incorrect one, which will entail a topology error. Topology errors usually cause the state estimation to be significantly biased, mainly because of the mismatch between measurements and system nodal admittance matrix. Apart from inaccurate results and convergence problems in state estimation, topology error may also cause bad data detection process to malfunction.

False data injection attack (FDIA) was first introduced in [2] and this class of cyber attack can mislead state estimation process by adding false data to measurements. Related work on FDIA can be roughly classified into the following categories. Unmitigated FDIA against SE [3]−[6], economic attacks on electricity market [7]−[9], detection and protection methods against FDIA [10], [11].

In this paper, we examine the possibility of false data injection which may misguide the system with incorrect topology. Without having to compromise the circuit breaker status, the adversary is shown to be capable of manipulating a topology error by adding false data injection to a number of analog measurements in the system. The attack is designed such that measurement residues will comply with the distribution of residues in the system with target topology error. Meanwhile, the difference between actual state values and theoretical state values is also minimized to match with topology error and avoid further detection based on state variables. The attack can mislead the system operators if the status of compromised branch is unknown or hacked. And after detection and identification of the designed topology error, adjusting system topology accordingly will lead the system to a biased SE result.

This paper is organized as follows. Section II introduces topology errors and existing detection methods. Section III illustrates the attack model towards topology errors. Simulation results of the proposed attack are given in Section IV and Section V concludes the paper.

## II. Topology Errors

This paper starts with the formulation of DC state estimation, although the proposed approach is generalizable towards AC state estimation as well. In DC state estimation, the relationship between measurements and state variables is linearized. Measurement model can be written as:

$$\boldsymbol{z} = \boldsymbol{H}\boldsymbol{x} + \boldsymbol{e} \qquad (1)$$

where $\boldsymbol{z}$ is real measurements, $\boldsymbol{x}$ is vector of bus angles and $\boldsymbol{e}$ is measurement error vector. $\boldsymbol{H}$ is the measurement Jacobian matrix.

The estimated state can be derived using Weighted Least Square (WLS) method:

$$\hat{x} = (H^T R^{-1} H)^{-1} H^T R^{-1} z \quad (2)$$

where $R^{-1}$ is diagonal weighting matrix.

And the residue can be given:

$$r = z - H\hat{x} \quad (3)$$

Based on (1) - (3), the following equations can be derived:

$$M = H(H^T R^{-1} H)^{-1} H^T R^{-1} \quad (4)$$

$$r = (I - M)e \quad (5)$$

Expected value of the residue after WLS state estimation:

$$E(r) = 0 \quad (6)$$

Errors in the status of breakers and switches will lead to incorrect information about network topology [12]. Topology errors are known to have more influence on measurement residues than parameter errors mainly because of the mismatch on $H$ matrix [13]. Let $H_t$ be the system true measurement Jacobian matrix and $H_e$ be the erroneous Jacobian matrix, then the error of the above Jacobian matrices which is caused by topology errors can be given as:

$$D = H_t - H_e \quad (7)$$

For a single branch status error, the non-zero elements in error Jacobian matrix $D$ are the first derivative of related measurements (injections on both buses and power flows on the branch) with respect to state variables (phase angles) of both buses. With known topology errors in the system, the measurement residue then becomes:

$$r = (I - M_e)(Dx + e) \quad (8)$$

$$M_e = H_e(H_e^T R^{-1} H_e)^{-1} H_e^T R^{-1} \quad (9)$$

And the expected value of the residue with topology error:

$$E(r) = (I - M_e)Dx \quad (10)$$

There exist certain types of topology errors that are non-detectable. Existing detection algorithms [12], [14], [15] commonly use measurement residue to detect and identify topology errors. A topology error is assumed to be detectable if $(I - M_e)Dx \neq 0$ for any state $x$. When a detectable topology error is present in the system, the bias vector can be represented as:

$$Dx = Lf \quad (11)$$

where $L$ is measurement to branch incidence matrix and $f$ is branch flow error vector.

The detection for a detectable topology error in the system can be dealt with using normalized residues [12]. From (11), it can be seen that the distribution of measurement residue of a system with topology error is related to both system topology and branch flow errors.

III. ATTACK MODEL

A. Preliminary

State estimation is usually solved as an overdetermined system, where there are more measurements than state variables and WLS method is commonly used to estimate the system state. Given a set of measurements $z$, the overdetermined system is able to obtain a unique estimated state $x$. Inversely, if given a fixed state $x$, the following linear system can be derived:

$$H^T R^{-1} z = (H^T R^{-1} H)\hat{x} = b \quad (12)$$

During state estimation, once $\hat{x}$ is determined and system topology is known then the right hand side of the above equation can be calculated. Consider $z$ as unknown variables, then the coefficient matrix for the system is $H^T R^{-1}$, which normally has more columns than rows. Furthermore, since the row vectors of system Jacobian matrix are linearly independent, $H^T R^{-1}$ is a full rank matrix. Suppose we have $m$ measurements and $n$ state variables, then for the above system:

$$rank(H^T R^{-1}) = rank(H^T R^{-1}|b) = n < m \quad (13)$$

The system becomes underdetermined and according to Rouché-Capelli theorem, there will be infinite number of solutions.

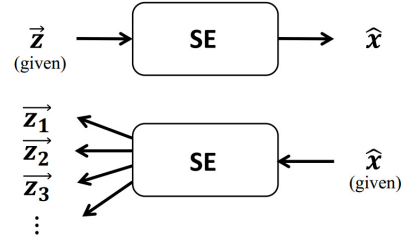

Fig. 1. Overdetermined and underdetermined systems

The underdetermined system as well as the parallelism in determining measurements provides possibilities of false data injection attacks in state estimation. The overdetermined and underdetermined systems are described in Fig. 1.

B. False data injection attacks

The goal of the proposed attack is to generate a topology error in the system by adding false data to system measurements. The attack will not compromise topology processor or circuit breakers, instead the attacker will utilize the redundancy in state estimation and manipulate measurement data to compensate the influence caused by the mismatch of Jacobian matrix through the topology error so that the system will appear to have a topology error.

Suppose the attacker knows the measurement data as well as system topology information a short time before the false data injection attack. Measurement data and Jacobian matrix at this stage can be denoted as $z$ and $H_t$. The real time measurement data taken right on/before the attack can be denoted as $z_t$. If the attacker is able to get the information about system right before the attack or can manipulate measurement data before it is obtained from the SCADA system, then $z$ can be treated as real time measurement data.

With measurement data $z_t$ and the topology information. The attacker is able to conduct state estimation based on the wrong topology. Given an intended topology, the measurement Jacobian matrix $H_e = H_t - D$ will be used in the state estimation:

$$\tilde{x} = (H_e^T R^{-1} H_e)^{-1} H_e^T R^{-1} z_t \quad (14)$$

The state above is the one when the system actually has a topology error and the attacker will expect that the state after the attack can be as close as possible to this state.

Assume the system topology does not change during the short time before the attack, the real time measurement Jacobian matrix can be known instantly since the non-zero elements in $H$ matrix are the line susceptance in the system. During real time state estimation, $H_t$ will be used as Jacobian matrix because the attacker only manipulate analog data but does not actually attack topology processor.

Therefore, for the real time state estimation the only unknown parameters are measurements since we already know $H$ matrix based on system topology and estimated state based on (14). The real time measurement data, which is the attacked one, can be represented as:

$$z_a = z_t + a \quad (15)$$

where $a$ is the false data vector on the measurement.

In most ideal situation, the adversary would expect the following equation to hold after the attack:

$$(H_t^T R^{-1} H_t)\tilde{x} = H_t^T R^{-1} z_a \quad (16)$$

After the false data injection, the attacker would want the system to identify it as a topology error rather than a malicious attack. The measurement residue distribution will change when the power flow values are compromised, but the attacker can easily find a legitimate measurement residue that is close to the residue in the ideal situation:

$$\|(z_a - H_t \tilde{x}) - (I - M_e) D \tilde{x}\|_\infty < \varepsilon \quad (17)$$

where $M_e = H_e(H_e^T R^{-1} H_e)^{-1} H_e^T R^{-1}$, $D = H_t - H_e$ and $\varepsilon$ is the threshold.

The state in (17) should be $(H_t^T R^{-1} H_t)^{-1} H_t^T R^{-1} z_a$ if (16) does not hold. Under the assumption that (16) holds or the difference between both sides is small enough, $\tilde{x}$ can be used as state vector in (17). The threshold can be adjusted and allow some violations for large systems.

The attacked measurement as well as the false data vector is bounded due to the constraint in (17). The determination of the entire set of measurement data not only needs to approach the estimated state in (14) after the state estimation but also needs to comply with the restrictions on the residue.

Suppose the system is free of attacks, then the real time state can be forecasted. State forecasting has been discussed in [10], [16], [17]. Autoregressive (AR) model can be used to forecast the state. The $i$th state variable on time $t$ can be expressed as:

$$x_t^i = \sum_{j=1}^{p}(\varphi_j^i x_{t-j}^i) + v_t^i \quad (18)$$

where $p$ is AR process order, $\varphi$ is AR parameters and $v_t$ is modeling uncertainties.

Parameter $\varphi$ can be represented with autocorrelations and solved using Yule-Walker equations. For each state variable, its AR parameter:

$$\begin{bmatrix}\varphi_1 \\ \varphi_2 \\ \vdots \\ \varphi_p\end{bmatrix} = \begin{bmatrix} 1 & \rho_1 & \cdots & \rho_{p-1} \\ \rho_1 & 1 & \cdots & \rho_{p-2} \\ \vdots & \vdots & \ddots & \vdots \\ \rho_{p-1} & \rho_{p-2} & \cdots & 1 \end{bmatrix}^{-1} \begin{bmatrix}\rho_1 \\ \rho_2 \\ \vdots \\ \rho_p\end{bmatrix} \quad (19)$$

where $\rho_k$ is the $k$th lag autocorrelation.

During the state forecasting, the dynamic model of system state is usually given as:

$$x_t = F_t x_{t-1} + v_t \quad (20)$$

where $F_t$ is the state transition matrix.

Note that $x_t$ above is different from the estimated state in (14) which is the state with topology error in the system. After the attack with system topology errors, the attacker would also want to see an obvious difference between the state after attack and the forecasted state, otherwise the attack itself is meaningless. The restriction on target state and the forecasted state is provided below, where $\delta$ is the threshold set before the attack. Note that if the attack data is prepared a short time before the attack for a detectable topology error, then (21) usually holds.

$$\|\tilde{x} - x_t\|_2 > \delta \quad (21)$$

Additionally, in real attack the adversary may only have access to limited number of measurements and the attack on each measurement may be restrained as well:

$$z_{a_{min}} \leq z_a \leq z_{a_{max}} \quad (22)$$

Therefore, based on all the discussion above, the attack can be formulated as an optimization problem:

$$\underset{z_a}{\text{minimize}} \quad \|(H_t^T R^{-1} H_t)\tilde{x} - H_t^T R^{-1} z_a\|_2$$

$$\text{subject to} \quad z_{a_{min}} \leq z_a \leq z_{a_{max}}$$

$$\|(z_a - H_t \tilde{x}) - (I - M_e) D \tilde{x}\|_\infty < \varepsilon$$

$$\|\tilde{x} - x_t\|_2 > \delta$$

where $z_{a_{min}}$ and $z_{a_{max}}$ are lower and upper bound of attacked measurements, the setup of which can be determined before the attack. For the measurement that the attacker is unable to compromise, fix the value of $z_{a_{min}}$ and $z_{a_{max}}$, after which the constraint of that measurement will turn into an equality constraint.

The attacker is able to obtain an optimal measurement data set for the intended attack on system topology with the above formulation. After the above optimization problem is solved, one can also easily get the attack vector $a = z_a - z_t$, which is also the false data to be injected to the measurement data.

## IV. ILLUSTRATIVE EXAMPLES

In this section, we illustrate the attack on the IEEE-14 bus system. A false data injection attack is performed with a forged topology error and an inclusion error is considered on branch 3-4. The branch is believed to be in service when it is actually open. The attacked 14-bus system is shown in Fig. 2.

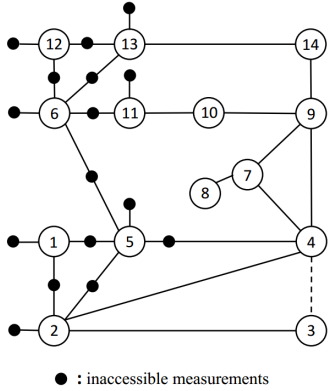

● : inaccessible measurements

Fig. 2. IEEE 14-bus system

Suppose the attacker is only able to compromise limited number of meters in the system during the attack. The measurements that cannot be attacked including injections and power flows are labelled in Fig. 2.

Assume the attacker can manage to obtain the measurement data a short time before the attack, the threshold $\varepsilon$ in (17) is set to be 0.8 in this example. Measurements considered are real injection and real power flows (including reverse power flows), and measurement set can be denoted as $z = [P_i, P_{ij}, P_{ji}]^T$. There are 54 measurement data in the system and after solving the proposed optimization problem, the attacker is able to get an optimal solution for $z_a$.

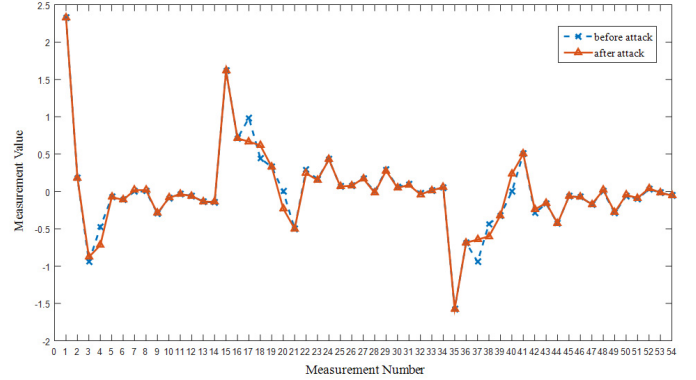

Fig. 3. Measurements before and after the attack

The measurements before and after the attack are shown in Fig. 3. The measurements that cannot be hacked as shown in Fig. 2 include power injections $P_i (z_k, k \in \{1,2,5,6,11,12,13\})$, real power flows $P_{ij}$ ($z_k, k \in \{15,16,19,21,24,25,26,27,33\}$) and power flows $P_{ji}$ ($z_k, k \in \{35,36,39,41,44,45,46,47,53\}$). These measurements are treated as inaccessible to the attacker, thus the value cannot be changed during the attack. The attack vector $a = z_a - z_t$ is shown in Fig. 4.

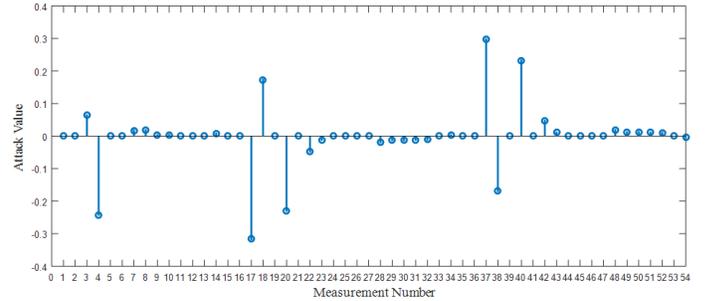

Fig. 4. Measurement attack vector

Since the measurements labelled in Fig. 2 are inaccessible to the attacker, the corresponding value of the attack vector should be zero, as shown in Fig. 4. After introducing the attack bias above, the system will have a residue that corresponds to a topology error on branch 3-4 as shown in Fig. 5, where the incident measurement residues of branch 3-4 are usually evident.

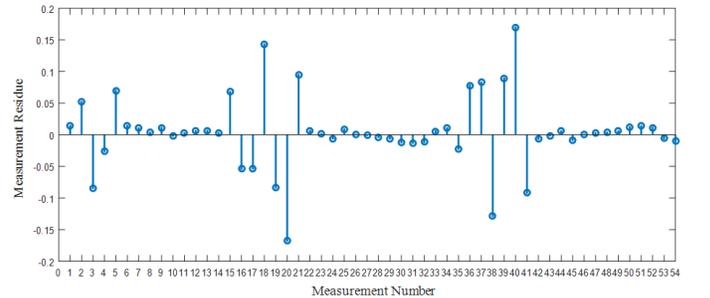

Fig. 5. Measurement residue after the attack

The objective function value of the attack model at solution is 0.0212. Basically the system state after the attack is equivalent to the theoretical state value $\tilde{x}$ in (14). The comparison of the state after attack and the theoretical one is shown in Fig. 6.

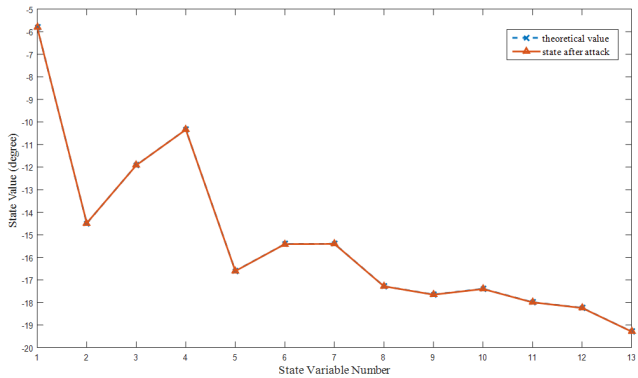

Fig. 6. Theoretical state and state after the attack

The difference of theoretical state value and actual state value at each bus (apart from slack bus) is small, as indicated in Fig. 7.

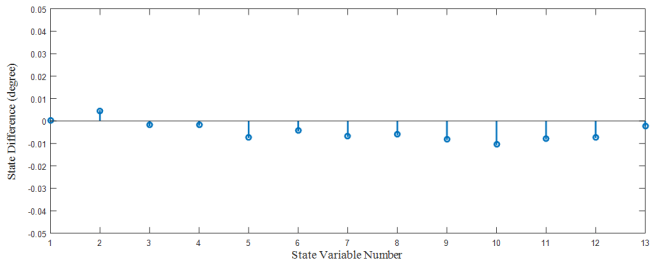

Fig. 7. Difference of state value at each bus

In this attack example, the absolute value of difference at each bus is below 0.01 degree. After the attack, it can be seen that the state values match perfectly with the theoretical values. Even with limited number of compromised measurements, the attacker is still able to forge a topology error in the system with a target state.

## V. CONCLUSION

In this paper, we propose a type of false analog data injection attack that can forge a topology error in the system. The attacker only needs to falsify analog measurement data to attack the system topology. Calculating measurement value given state values is an underdetermined problem, therefore the adversary is able to generate multiple attack bias as a result of system redundancy. The attack vector can be obtained by solving an optimization problem, after which the state value should converge to the theoretical state value. A theoretical state value is computed at the beginning based on intended topology and then the attacker will conduct a reverse estimation on measurement values through an optimization model. Even with limited number of target measurements, the attacker is still able to finish an attack towards system topology errors.

The proposed attack on topology is detrimental to the system since the manipulation on analog measurement data by the attacker would mislead the system operators to identify it as topology errors rather than unintended measurement bad data. And changing the branch status after detecting such errors will lead to a different system state result.

The feasibility of the proposed false data injection attack is tested on IEEE 14-bus system. The attacker can expect better results if the number of inaccessible measurements is smaller and the system is larger, as it will bring more redundancy to the attack model. Future work may include detection and identification of such false data injection attacks.